\documentstyle[11pt]{article}
\date{}

\begin{document}

\setcounter{page}{0}

\begin{flushright}
CERN-TH. 95-34
\end{flushright}
\vspace{1.0cm}

\begin{center}
{\LARGE
{\bf REAL-TIME DYNAMICS OF PARTON-HADRON CONVERSION}

}
\end{center}
\bigskip

\begin{center}

{\Large
{\bf J. Ellis and K. Geiger}
}

{\it CERN TH-Division, CH-1211 Geneva 23, Switzerland}
\end{center}
\vspace{2.0cm}

\begin{center}
{\large {\bf Abstract}}
\end{center}
\medskip

\noindent
We propose a new and universal approach to the hadronization problem
that incorporates both perturbative QCD and effective field theory
in their respective domains of validity and that models the transition
between them in analogy to the finite temperature QCD phase transition.
Using techniques of quantum kinetic theory, we formulate a
real-time description in momentum and position space.
The approach is applied to the evolution of  fragmenting $q\bar q$ and $gg$
jets
as the system evolves from the initial 2-jet, via parton multiplication and
cluster formation, to the final yield of hadrons.
We investigate time scale of the transition, energy dependence, cluster
size and mass distributions, and compare our results for particle production
and Bose-Einstein correlations with experimental data
for $e^+e^-\rightarrow hadrons$.  An interesting possibility to extract
the space-time evolution of the system from Bose enhancement measurements
is suggested.
\noindent

\vspace{0.5cm}

%\leftline{PACS Indices: .....................}
\rightline{johne@cernvm.cern.ch}
\rightline{klaus@surya11.cern.ch}
\leftline{CERN-TH. 95-34, March 1995}

\newpage

The physics of QCD is well understood in two limits:
at short distances ($\ll$ 1 fm) where the relevant
degrees of freedom are quarks and gluons,
the dynamics of which is accurately described by perturbative QCD,
and at large distances ($\,\lower3pt\hbox{$\buildrel > \over\sim$}\,1$ fm)
where the relevant degrees of freedom are hadrons
whose non-perturbative interactions are well described by
chiral models. What is less understood, and constitutes one
of the key open problems in QCD, is the transition between the
short- and long-distance regimes through intermediate distance scales.
This problem is particularly serious for
attempts to describe the phenomenon
of hadronization in high energy QCD processes.
Perturbative QCD describes quantitatively the short-distance
evolution of the dynamical system and hence the large-scale flow
of energy-momentum. The transformation of
colored quarks and gluons into colorless hadrons,
however, is commonly
believed not to involve large momentum transfers, as expected for
a large-distance effect. Moreover, experiments on many
high-energy QCD processes, including for example $e^+e^-\rightarrow hadrons$
and deep inelastic lepton-nucleon scattering, strongly support the
idea that the fragmentation of partons into hadrons is a universal
mechanism.
However, the theoretical tools currently
available for studying QCD are inadequate to describe the dynamics of this
transformation from partonic to hadronic degrees of freedom. Perturbative
techniques are limited to the short distance regime where confinement
is not apparent \cite{dok80}, whilst effective low-energy chiral models
\cite{chiral}
and QCD sum rules \cite{shifman79} that incorporate confinement, lack partonic
degrees of
freedom.
In principle, lattice QCD \cite{karsch89} should be able to bridge the gap, but
in practice
dynamical calculations of parton-hadron conversion are not yet feasable.

In this paper we extend previous work \cite{ms36} and
advocate a new approach to the hadronization problem
that incorporates both perturbative QCD and effective field theory
in their respective domains of validity and that models the transition
between them using ideas developed in phenomenological
descriptions of the finite temperature transitions
from a quark-gluon plasma to a hadronic phase \cite{CEO}.
The latter is described by an effective theory that incorporates a chiral field
$U$ whose vacuum expectation value ($vev$)
$U_0 \equiv \langle 0 |U+U^\dagger|0\rangle$
represents the quark condensate $\langle 0 |\bar q q|0\rangle$,
and a scalar field $\chi$ whose $vev$ $\chi_0 \equiv \langle 0 | \chi| 0
\rangle$
represents the gluon condensate
$\langle 0 | F_{\mu\nu}F^{\mu\nu}| 0 \rangle$.
We visualize a high energy collision such as $e^+e^- \rightarrow \bar q \,q$
as producing a "hot spot" in which the long range order,
represented by $U_0$ and $\chi_0$, is disrupted locally by the
appearance of a bubble of the naive perturbative
vacuum in which
$\langle 0 |\bar q q|0\rangle = 0=\langle 0 | F_{\mu\nu}F^{\mu\nu}| 0 \rangle$.
Within this bubble, a parton shower develops
in the usual perturbative way, with the hot spot expanding
and cooling in an irregular stochastic manner described by QCD transport
equations.
This perturbative description remains appropriate in any
phase-space region of the shower where the local energy density is large
compared with the difference in energy density between
the perturbative partonic and the non-perturbative hadronic vacua.
When this condition is no longer satisfied, a bubble of hadronic
vacuum may be formed with a probability determined by statistical-mechanical
considerations. A complete description of this conversion
requires a treatment combining partonic and hadronic degress of freedom,
which is the essential aspect of our approach.
Although studies of the
finite temperature QCD phase transition \cite{kajantie94}
indicate that it may be completed quite rapidly, in which case mixed
description is not needed in a first approximate treatment,
we are more ambitious here and propose a universal approach to the dynamic
transition {\it between} partons and hadrons based on an effective
QCD field theory description and relativistic kinetic theory that includes
a mixture of both sets of degrees of freedom.
\smallskip

The purpose of this letter is to present the essential concepts of our
approach,
i.e. the quantitative formulation and its application of the above picture.
An extensive documentation of our work can be found in Ref. \cite{ms37}
to which we refer for details.
Let us begin by defining the {\it distance measure} $L$ for
the space-time separation between two points $r$ and $r'$
[$r\equiv r^\mu=(t,\vec r)]$:
\begin{equation}
L\;:=\;\sqrt{(r-r')_\mu(r-r')^\mu}
\;,
\label{Ldef}
\end{equation}
and introduce a {\it characteristic length scale} $L_c$ that separates
short distance ($L\ll L_c$)
and long range ($L\,\lower3pt\hbox{$\buildrel > \over\sim$}\,L_c$)
physics in  QCD. The scale $L_c$ can be associated with the
confinement length of the order of a hadron radius.
In the limit of short distances $L\ll L_c$ (or high momenta, high
temperatures),
QCD may be described perturbatively by a Lagrangian
in terms of the {\it elementary} gluon ($A^\mu$) and quark fields
($\psi,\overline{\psi}$),
\begin{equation}
{\cal L}_L[A^\mu,\psi,\overline{\psi}]
\;=\;
-\frac{\kappa_L}{4}\,\,F_{\mu\nu, a} F^{\mu\nu}_a
\;+\;  \overline{\psi}_i \left[\frac{}{}\,\left(\frac{}{}i \gamma_\mu \partial
^\mu
- \mu_L\right) \delta_{ij}
- g_s  \gamma_\mu A^\mu_a T_a^{ij} \right]\, \psi_j
\;,
\label{L1}
\end{equation}
where
$F_a^{\mu\nu}= \partial^\mu A_a^\nu -\partial^\nu A_a^\mu + g_s f_{abc} A^\mu_b
A^\nu_c$
and summation over color indices $a$, $i,j$ is understood.
The functions $\kappa_L$ and $\mu_L$
introduce an explicit  scale($L$)-dependence in ${\cal L}_L$ which modifies the
quark and
gluon properties
when $L$ increases towards $L_c$ and beyond. In the limit $L\rightarrow 0$,
$\kappa_L=1$ and $\mu_L=0$ (neglecting the quark current masses),
thus the QCD Lagrangian is recovered.
However, at larger $L$,  the bare quark and gluon fields become
dressed by non-perturbative dynamics and we expect $\kappa_L < 1$ and $\mu_L\ne
0$.
In fact, the short range behaviour of $\kappa_L$ (and similarly of $\mu_L$)
can be calculated perturbatively:
$\kappa_L = [1+ g_s^2/(8\pi)^2 (11-2 n_f/3) \ln(L\,\Lambda)]^{-1}$.
Hence both $\kappa_L$ and the dynamical quark masses $\mu_L$ vary
(as does the coupling constant $g_s$) with the renormalization
scale $\Lambda$, which we expect to be directly related to the confinement
scale $L_c$.
We will return to that issue below.

In the limit of large distances
$L\,\lower3pt\hbox{$\buildrel > \over\sim$}\,L_c$ (or low momenta, low
temperatures)
the hadronic physics can well be described by an effective field theory
that embodies the scale and chiral constraints of the fundamental QCD
Lagrangian.
The corresponding effective Lagrangian is written in terms of
{\it collective} fields $\chi$, $U,U^\dagger$ \cite{CEO}:
\begin{equation}
{\cal L}[\chi,U,U^\dagger]
\;=\;
\frac{1}{2}\,(\partial_\mu \chi) ( \partial^\mu \chi )
\;+\; \frac{1}{4}\,
Tr\left[\frac{}{}(\partial_\mu U)
( \partial^\mu U^\dagger )
\right]
\;-\;V(\chi,U)
\;,
\label{L2}
\end{equation}
where $\chi$ is the aforementioned scalar field and
$U = f_\pi \exp\left(i \sum_{j=0}^8 \lambda_j\phi_j/f_\pi\right)$
the  pseudoscalar field for the nonet of  meson fields $\phi_j$,
and the potential $V(\chi,U)$ is given by
\begin{eqnarray}
V(\chi, U) &=&
b \;\left[ \frac{1}{4} \,\chi_0^4 \;+\;
\chi^4 \,\ln\left(\frac{\chi}{e^{1/4} \chi_0}\right)\right]
\;+\;
\frac{1}{4}\,\left[1\,-\,\left(\frac{\chi}{\chi_0}\right)^2\right] \;
Tr \left[ (\partial_\mu U)(\partial^\mu U^\dagger)\right]
\nonumber \\
& & +\;
 c\; Tr\left[\frac{}{} \hat m_q (U + U^\dagger) \right]
 \; \left(\frac{\chi}{\chi_0}\right)^3
\;,
\label{V}
\end{eqnarray}
which models  the quantum dynamics in the low energy regime associated with the
self-
and mutual interactions of $\chi$ and  $U$.
Here the parameter $b$ is related to the conventional bag constant $B$ by
$B=b\chi_0^4/4$,
$c$ is a constant of mass dimension 3,  $m_q = \mbox{diag}(m_u,m_d,m_s)$
is the light quark mass matrix.
The $vev$'s of the collective fields $\chi$ and $U$ are zero in the
short distance limit $L\ll L_c$, i.e. in the naive perturbative vacuum, but
become finite in the long wavelength limit
at large distances $L\,\lower3pt\hbox{$\buildrel > \over\sim$}\,L_c$,
namely $\langle0| \chi |0\rangle = \chi_0$ and $\langle 0| U+U^\dagger|0
\rangle = U_0$,
which are to be regarded as order parameters of the physical vacuum.
Thus, the potential (\ref{V})
has a  minimum when $\langle \chi\rangle =\chi_0$
and equals the vacuum pressure (bag constant) $B$ at $\langle \chi\rangle =0$.

As already advertised, we seek to describe parton-hadron conversion in the
aftermath
of a high-energy collision, in which an initial ``hot spot'' expands
irregularly. Inhomogenities that appear during the expansion are characterized
by
local dynamical scales. It is well known how to incorporate small distance
(high momentum) scales using the renormalization group in perturbative QCD.
Dynamical scale dependence can also be taken into account in the long distance
effective Lagrangian (\ref{L2}).
For instance, at finite temperature $T$ the effective potential
(\ref{V})
acquires additional  $T$-dependent terms \cite{CEO}, resulting in a
modification
of the $vev$'s of $\chi$ and $U+U^\dagger$ and thus of the gluon and
quark condensdates,
$\langle 0 | F_{\mu\nu}F^{\mu\nu}| 0 \rangle$ and
$\langle 0 |\bar q q|0\rangle$.
In the present situation we are dealing however with general non-equilibrium
systems, in which we advocate the length scale $L$ to control
the scale dependence of local fluctuations.
A complete treatment of the hadronization process clearly requires
the linkage between the elementary partonic degrees of freedom ($A^\mu, \psi$)
and the collective hadronic degrees of freedom ($\chi, U$). From the above
discussion we expect that
\begin{eqnarray}
\lim_{L\rightarrow 0} \kappa_L \,=\, 1
\;,& &\;
\lim_{L\rightarrow 0} \mu_L \,=\, 0
\;,\;\;\;\;\;\;\;\;
\lim_{L\rightarrow 0} \langle \chi\rangle_L \,=\, 0
\nonumber \\
\lim_{L\rightarrow \infty} \kappa_L \,=\, 0
\;,& &\;
\lim_{L\rightarrow \infty} \mu_L \,=\, \infty
\;,\;\;\;\;
\lim_{L\rightarrow \infty} \langle \chi\rangle_L \,=\, \chi_0
\label{limits}
\end{eqnarray}
corresponding to the unconfined and confined phases, respectively.
In view of the one-to-one relation between $L$ and $\langle \chi\rangle_L$
one may express (\ref{limits}) as boundary conditions of the
$\langle \chi\rangle$-dependence of $\kappa_L$ and $\mu_L$ when
$\langle \chi\rangle\rightarrow 0$ or
$\langle \chi\rangle\rightarrow \chi_0$.
This characterizes the so-called color-dielectric property of the
vacuum in QCD \cite{FL}, which has been widely
applied in soliton models and hadron phenomenology.

It is important to realize that the variation
of the internal length scale $L$, as defined in (\ref{Ldef}),
is governed by the dynamics of the fields itself and
in turn it must determine the time evolution of the
fields $A^\mu, \psi$ as well as $\chi$.
With regard to (\ref{limits}), we assume that
the behavior of the functions $\kappa_L$ and $\mu_L$ is correlated with the
value of
$\langle \chi \rangle$ and express this by
$\kappa_L\equiv\kappa_L(\chi)$ and $\mu_L\equiv\mu_L(\chi)$.
Then, by combining (\ref{L1})-(\ref{V}), we obtain
an effective field theory description covering
the full range $0 < L < \infty$,
\begin{equation}
{\cal L}_L[A^\mu,\psi,\chi,U]
\;:=\; {\cal L}_L[A^\mu,\psi,\overline{\psi}]
\;+\;{\cal L}[\chi,U,U^\dagger]
\;,
\label{genf4}
\end{equation}
in which the elementary gluon and quark fields are coupled to the
collective field $\chi$ via
$\kappa_L(\chi)$ and $\mu_L(\chi)$ in a self-consistent manner
that is controlled by the dynamically varying scale $L$.
Our present understanding of QCD does not yield us explicit
interpolating formulae for $\kappa_L$ and $\mu_L$ from first principles.
Fortunately, as we find, the particularities of their functional forms are not
crucial for the treatment of parton hadron conversion that we present here,
as long as they are smooth and satisfy the boundary conditions (\ref{limits}).
The reason is that, as $L$ decreases, the appropriate dynamical
description of QCD apparently jumps rapidly from the first limit in
(\ref{limits}) to the neighborhood of the second limit, corresponding
to the rapid formation of hadronic domains when
$L\,\lower3pt\hbox{$\buildrel > \over\sim$}\, L_c$. This feature
is very reminiscent of the weakly first order nature of the QCD phase
transition
at finite temperature.
The essential point is that, due to the conditions (\ref{limits}),
the variation of $\kappa_L$ with $L$ and $\chi$  generates
color charge confinement at large distances, because of the fact that
a color electric charge creates a displacement $\vec D_a = \kappa_L \vec E_a$,
where $E_a^k = F_a^{0k}$, with energy $\frac{1}{2} \int d^3 r D_a^2/\kappa_L$
which becomes infinite at large $L$ for non-zero total charge.
Similarly, the increase of the dynamical mass $\mu_L$ with $L$ and $\chi$
leads to an effective confinement potential that
ensures absolute confinement also for quarks by prohibiting their
propagation at large distances.
We tested various choices for $\kappa_L$ and $\mu_L$ and,
to be specific, in what follows we adopt the forms
$\kappa_L(\chi)= 1- (L\chi)^2/(L_0\chi_0)^2$, which is
a minimal possibility compatible with the QCD scaling properties, and
$\mu_L(\chi)= \mu_0(\kappa_L^{-1}(\chi)-1)$ with constant $\mu_0$ set equal to
1 GeV.
The latter form reflects that the quark mass term $\mu_L$ is
assumed to be induced by non-perturbative gluon interactions (through
$\kappa_L$),
rather than being an
independent quantity, as is suggested by an explicit calculation \cite{CDM}
of the quark self-energy involving the gluon propagator in the
presence of the collective field $\chi$.
\smallskip

Let us emphasize once more the similarity of the above approach with
finite temperature QCD phenomenology \cite{CEO}.
The effect of $\kappa_L$ and $\mu_L$  can be interpreted as a scale ($L$)
dependent modification $\delta V$, which adds to the ($L$ independent)
potential $V$, eq. (\ref{V}),
\begin{equation}
{\cal V}(L) \;:=\;
V(\chi, U) \;+\;\; \delta V(L,\chi)
\;\;,\;\;\;\;
\delta V(L,\chi) \;=\;
\frac{\kappa_L(\chi)}{4} \; F_{\mu\nu, a} F^{\mu\nu}_a
\;+\;
\mu_L(\chi) \;\overline{\psi}_i\, \psi_i
\;,
\label{calV}
\end{equation}
with  $\delta V = O(L^{2})$.
Here the length scale $L$ corresponds to the temperature $T$
in finite temperature QCD, where the correction to the zero temperature
potential is $O(T^2)$.
Hence we expect the characteristic confinement length $L_c$ to play a similar
role as the
critical temperature $T_c$ in the QCD phase transition.
However, this formal anology is to be taken with some caution, because
here we are aiming to describe the evolution of a general non-equilibrium
system in real time and Minkowski space, as opposed to the thermal evolution
in Euclidian space.

We complete our discussion of the above field theory aspects with
the following important remarks:
(i) The formulation is gauge and Lorentz invariant and is consistent
with scale and chiral symmetry properties of QCD. It interpolates from
the high momentum (short distance) QCD phase with unconfined gluon
and quark degrees of freedom and chiral symmetry
($\langle \chi\rangle=0, \langle U \rangle =0, \kappa_L =1, \mu_L=0$), to a
low energy (long range) QCD  phase with confinement and broken chiral symmetry
($\langle \chi\rangle=\chi_0, \langle U \rangle =U_0, \kappa_L =0,
\mu_L=\infty$).
(ii) By construction  (\ref{genf4}) strictly avoids double counting of
degrees of freedom, because the introduction of the scale $L$ and the
behaviour of the $L$-dependent the functions $\kappa_L$ and $\mu_L$
truncate the dynamics of the elementary fields $A^\mu$, $\psi$
to the short distance regime, whereas the effective description
in terms of the collective fields $\chi$, $U$ covers the complementary
long range domain.
(iii) There is no need for explicit renormalization of
the composite fields  $\chi$ and $U$, because they  are already interpreted as
effective
long range degrees of freedom with loop corrections implicitely included in
$V(\chi,U)$
and it would be double counting to add them again.
\medskip

As documented in detail in \cite{ms37,ms39}, it is possible to derive from
the Lagrangian (\ref{genf4})
a formulation of QCD transport theory that incorporates
both partons and hadrons, yielding a fully dynamical description of
the QCD matter in real time and complete phase-space.
Since this serves as the basis for our subsequent description of
the conversion of partons into hadrons locally in phase-space, we sketch
here  this formulation of QCD transport theory.
The first step is to derive from the field equations of motion
the corresponding Dyson-Schwinger equations for the {\it real-time Greens
fuctions}
of the fields $\psi$, $A^\mu$, $\chi$, and $U$, denoted by
$G_\alpha(x,y)$, and
where $\alpha\equiv$ $q$, $g$, $\chi$, $U$.
They are
defined as the two-point functions that measure the time ordered correlations
between the fields at space-time points $x$ and $y$. In symbolic
notation, one obtains a coupled system of integral equations,
\begin{equation}
G_\alpha (x,y)\;=\; G_\alpha^{(0)}(x,y) \;+\;
\sum_\beta \,\int d^4x' d^4x''\,G_\alpha^{(0)}(x,x') \,\Sigma_\beta(x',x'')\,
G_\alpha(x'',y)
\;,
\label{eog4}
\end{equation}
where $G_\alpha^{(0)}$ denotes the free field Greens functions that
satisfy the equations of motion in absence of interactions, and
the  {\it self-energies} $\Sigma_\alpha$
embody both the mutual and the self-interactions of the fields.
The explicit expressions can be found in Ref. \cite{ms39}.
Next we introduce the Wigner transforms
\begin{equation}
{\cal G}_\alpha(r,p)\;=\; \int d^4 R \,e^{ip\cdot R} \,G_\alpha(x,y)
\label{wigner}
\;,
\end{equation}
(where $r=\frac{1}{2}(x+y)$, and $p$ is the conjugate variable to $R=x-y$)
of the Greens functions and similarly for the self-energies
$\Sigma_\alpha(x,y)$.
The dependences on $r$ would be trivial for field theory {\it in vacuo},
but non-trivial in an inhomogenous QCD matter.
In terms of the Wigner transforms ${\cal G}_\alpha(r,p)$ and
$\tilde{\Sigma}_\alpha(r,p)$,
the Dyson-Schwinger equations (\ref{eog4})
become kinetic  equations describing the real-time evolution
of the matter in phase-space spanned
by $\vec r$ and $p^\mu=(E,\vec p)$.
Then, by tracing over color and spin polarizations, and taking
the expectation values (or, in medium the
ensemble average) of the Wigner transformed Greens functions
${\cal G}_\alpha$,
one obtains the scalar functions
\begin{equation}
 F_\alpha(r,p)\;\,\equiv\; \, F_\alpha (t, \vec r; \vec p, p^2=M_\alpha^2)
 \;=\; \langle \; Tr[ {\cal G}_\alpha(r,p) ] \;\rangle
\left.\frac{}{} \right|_{M_\alpha^2=\tilde{\Sigma}_\alpha(r,p)}
\;.
\label {WWWW}
\end{equation}
The $c$-number functions $F_\alpha(r,p)$ for the
particle species $\alpha$
are the quantum mechanical analogues to the
classical phase-space distributions that measure the number of particles
at time $t$ in a  phase-space element $d^3rd^4p$.
The $F_\alpha$ contain the essential  microscopic information required for a
statistical description of the time-evolution of a many-particle system in
complete phase-space and provide the basis for calculating
macroscopic observables
in the framework of relativistic kinetic theory.
In particular, the local space-time dependent
particle currents $n_\alpha$ for the different particle species
and the corresponding energy-momentum tensors
$T_\alpha^{\mu \nu}$ are given by \cite{msrep}
\begin{equation}
n_\alpha^\mu (r) \;=\;
\int d\Omega_\alpha \, p^\mu F_\alpha (p,r)
\;,\;\;\;\;\;\;\;
T_\alpha^{\mu \nu} (r)
\;=\;
\int d\Omega_\alpha p^\mu p^\nu \, F_\alpha (r, p)
\label{Tmunu}
\;,
\end{equation}
where
$d\Omega_\alpha=\gamma_\alpha  dM^2 d^3 p/(16\pi^3p^0)$,
the $\gamma_\alpha$ are degeneracy factors for the internal
degrees of freedom (color, spin, etc.),
$M_\alpha$ measures the amount by which a particle $\alpha$ is off mass-shell
as a result of the selfenergy terms in (\ref{eog4}),
and $p^0\equiv E= +\sqrt{\vec p^{\,2} + M_\alpha^2}$.
These macroscopic quantities can be written in  Lorentz invariant form
by introducing for each species $\alpha$ the associated  matter flow velocity
$u_\alpha^\mu(r)$,
defined as a unit-norm time-like vector at each space-time point,
$(u_\mu u^\mu)_\alpha = 1$. A natural choice is
e.g. $u_\alpha^\mu=n_\alpha^\mu /\sqrt{ n_{\nu\,\alpha} n_\alpha^\nu}$.
By contracting the quantities (\ref{Tmunu}) with the
local flow velocities $u_\alpha^\mu$, one can now obtain
corresponding invariant scalars of  particle density,
pressure, and energy density, for each particle species $\alpha$ individually,
\begin{equation}
n_\alpha (r)\;=\; n_{\mu\,\alpha}\, u_\alpha^\mu
\;\;,\;\;\;\;
P_\alpha (r) \;=\; - \frac{1}{3} \,T_{\mu \nu,\,\alpha} \,
\left( g^{\mu \nu}-  u_\alpha^\mu u_\alpha^\nu \right)
\;\;,\;\;\;\;
\varepsilon (r) \;=\; T_{\mu \nu ,\,\alpha}\,  u_\alpha^\mu \, u_\alpha^\nu
\label{pr}
\;.
\end{equation}

The above formulation is readily applicable to the dynamics of the
parton-hadron conversion in rather general situations.
Here we will as an illustrative application study
the fragmentation of a $q\bar q$ jet system
with its emitted bremsstrahlung gluons and describe the evolution of the
system as it converts from the parton phase to the hadronic phase.
In this  case
the kinetic equations that derive from the Dyson-Schwinger equations
(\ref{eog4}) simplify considerably and yield a set of coupled tranport
equations
for the phase-space densities
$F_\alpha (t,\vec r;\vec p, M_\alpha^2)$  of the generic form \cite{ms37}
\begin{equation}
\left[
p_\mu \,\partial_r^\mu
\;+\; (\overline{M}_\alpha \,\partial_r^\mu \overline{M}_\alpha)
\,\partial_\mu^p
\right] \; F_\alpha
\;=\;
\sum_{processes\; k} \left[\hat I_k^{(+)}(F_\beta)\;
-\;\hat I_k^{(-)}(F_\beta)\right]\;  F_\alpha
\label{Feq}
\;.
\end{equation}
These kinetic equations reflect a probabilistic
interpretation of the evolution in terms of successive
interaction processes $k$,
in which the change of the particle distributions $F_\alpha$
is governed by the balance of gain (+) and loss ($-$) terms.
The left hand left side describes  propagation of a
quantum of species $\alpha$ in the presence of the
mean field $\overline{M}_\alpha$ generated by the others, and on the
right hand side the integral operators $\hat I^{(\pm)}$
incorporate the effects of the self-energies in terms of
real and virtual interactions that lead to particle excitations.
\smallskip

We now discuss the illustrative application of the outlined formalism to
parton-hadron conversion starting from
a highly virtual $\gamma^\ast$ or a $Z^0$ in an $e^+e^-$ annihilation event
with large invariant mass $Q^2 \gg \Lambda^2$
corresponding to a very small initial $L\ll  1\;fm$, that initiates
a  $q \bar q$ jet evolution.
At first the partons multiply, determined by a
coherent parton shower simulation incorporating (coherent) gluon emission,
secondary $q\bar q$ production, etc..
Eventually, as the quanta diffuse in space-time and $L$ increases,
they will coalesce to pre-hadronic clusters by tunneling through the
potential barrier imposed by the
dynamically changing  ${\cal V}(L)$, which is determined
within a coalescence model.
Finally these pre-hadronic clusters will  convert
into physical hadronic states and subsequently decay into low mass hadrons,
depending on the density of accessible
hadronic states.  The system of particles is evolved in discrete time steps,
here taken
as $\Delta t = 0.01$ $fm$, in coarse grained 7-dimensional phase-space with
cells $\Delta \Omega = \Delta^3 r \Delta^3 p \Delta M^2$.
The partons propagate along classical trajectories until they interact,
i.e. decay (branchings) or recombine (cluster formation). Similarly, the
formed clusters travel along straight lines until they decay into hadrons.
The corresponding probabilities and time scales of interactions are
sampled stochastically from the relevant probability distributions.
With this concept, we can trace the space-time evolution of the
system \cite{ms37}:
In each time step, any ``hot'' off-shell parton is allowed
to decay into ``cooler'' partons, with a probability determined
by its virtuality and life-time. Also in each step, every parton and its
nearest
spatial neighbor are considered as defining a fictious pre-hadronic bubble
with invariant radius $L$, as defined by (\ref{Ldef}).
Depending on the value of $L$, the relative probabilities for which
configuration is more favorable
determine whether the partons continue in their shower development, or a parton
cluster is formed.
This cascade evolution is followed until all partons
have converted, and all clusters have decayed into final hadrons.

To this end we need to specify the parameters in the
potential $V$, eq. (\ref{V}), or ${\cal V}(L)$, eq. (\ref{calV}).
As $L$ varies, ${\cal V}$ changes its shape and affects the
evolution of the system.
The details of the dynamics are controlled the choice of
bag constant $B$ which defines the
vacuum pressure $V(0)$ in the short distance limit $L\rightarrow 0$,
and $\chi_0$, the value of the condensation of $\chi$ in the long
distance regime.
Although the values of $B$ and $\chi_0$ are not precisely known, there
is agreement of various phenomenological determinations about their range:
one expects \cite{CEO} $B^{1/4}=(150 - 250)$ MeV and
$\chi_0=(50 - 200)$ MeV. In what follows, we choose
two representative combinations:
$(B^{1/4},\chi_0)=(230, 200)$ MeV
and $(B^{1/4},\chi_0)=(180, 100)$ MeV.
We then proceed in analogy to the QCD phase transition model of Ref. \cite{CEO}
and compare in each time step the local pressure of partons $P_{qg}(t,\vec
r,L)$ with the
pressure $P_{\chi}(t,\vec r,L)$  that a pre-hadronic bubble would
create instead. We compute the pressures from the corresponding
phase-space densities (\ref{WWWW}) and the formulae (\ref{Tmunu}),(\ref{pr}).
Representing
\begin{equation}
P_{qg}(r,L) \;=\; a_{qg}(r,L)\;L^{-4} \;-\; B
\;\;,\;\;\;\;\;
P_{\chi}(r,L) \;=\;a_\chi(r,L)\;L^{-4}\;-\; {\cal V}(L)
\;,
\label{P}
\end{equation}
and, defining $L=L_c$ as the characteristic length scale, in analogy
to the critical temperature in the QCD phase transition, such that the
two pressures equal each other at $L_c$ and ${\cal V}(\chi,L_c)=0$, we get the
determining condition:
\begin{equation}
L_c\;=\; \left[\frac{a_{qg}(r,L_c)\;-\;a_\chi(r,L_c)}{B}\right]^{1/4}
\label{Bval}
\;.
\end{equation}
With the functions $a_{qg}$ and $a_\chi$  obtained from the
numerical simulation, we find then for the characteristic
confinement length
$L_c = 0.6$ $fm$ for the choice
$B^{1/4}=230$ MeV and $L_c = 0.8$ $fm$ for $B^{1/4}=180$ MeV.

Fig. 1a compares the parton and hadron pressures $P(t,L)$
(calculated at the peak of the expanding shock front) as a function of
time $t$ in parton showers initiated by $q\bar q$-pairs with
center-of-mass energies $Q=100$ GeV.
We see that the pressures cross over at a time $t\simeq$ 0.7 $fm$, almost
independent of the choice of $B$ (and hence $L_c$), and we have found
the same feature in $gg$-initiated showers.
The corresponding time development of the total generated transverse momentum
$p_\perp(t)$ is shown in Fig. 1b, where we see that
the share of hadronic clusters dominates when
$t \,\lower3pt\hbox{$\buildrel > \over\sim$}\, 1$ $fm$,
which again is also true for $gg$-initiated showers.
At lower energy
(we performed an identical analysis at $Q=10$ GeV)
we found the same features, except that the crossover times are
somewhat shorter.

In this dynamically evolving situation the probability that
two partons with invariant spatial separation $L$ coalesce to
a cluster (viewed as a pre-hadronic bubble formed in the vacuum),
is determined by $\pi(L)=1-\exp(-\Delta F \,L)$, where $\Delta F$ is the free
energy of a thin-walled bubble of one phase (partons), immersed in a
medium in the other phase (vacuum):
\begin{equation}
\Delta F\;=\; \frac{4\pi}{3}\; R_c^2\;\sigma
\;,
\label{k3}
\end{equation}
where $R_c=2\sigma/\Delta P$ is the critical bubble size
determined in terms of the surface tension $\sigma$
and the difference in pressures $\Delta P = P_{qg}- P_\chi$. We estimate
the surface tension from the relation \cite{coleman}
$\sigma \equiv \sigma(L_c)=\int d\chi \sqrt{2 {\cal V}(L)} |_{L=L_c}$
with the potential (\ref{calV}),
which yields $\sigma^{1/3}= 40$ (48) MeV (for $B^{1/4} = 230$ (180) MeV).
We remark that similar numbers are indicated by lattice QCD simulations.
These small values of $\sigma$ correspond to a weakly
first-order transition at finite temperature, which is consistent
with astrophysical constraints
on inhomogenities. As a consequence of this feature,
although our approach allows for parton-hadron
conversion when the parton separation $L$
is even less than $L_c$, or, when $L\gg L_c$, most clusters in
fact form when $L$ is only slightly larger than $L_c$,
as seen in Fig. 2a.
We also observe that the distribution of cluster sizes is essentially
independent of the
initial state energy. The same feature was found true for
$gg$-initiated showers.
The spectrum of cluster masses, determined by the total invariant
masses of coalescing partons at the moment of conversion, is shown
in Fig. 2b which exhibits no sensitivity to the value of $L_c$ at all.
Moreover, the shapes of both spectra turn out to be rather independent of the
jet energy, as well as of the type of initial partons,
and therefore appear to be universal.

To obtain the hadron spectrom from the cluster distribution,
we implement essentially the cluster-fragmentation scheme of
\cite{pcm}, but with the important modification that a cluster
weighing more than a critical mass $M_{crit}=4 GeV$ is not
allowed to convert from partons to hadrons, but is forced to continue
its partonic shower development. This restriction
is motivated by the experimental indication that
very heavy cluster production is suppressed.
Although conceptionally significant,
it is not very important numerically, because
it affects less than 5 \% of the clusters.
We show in Fig. 3a the total charged particle multiplicity
$n_{ch}$, which grows with energy in consistency with
experimental data.
The momentum spectra of charged hadrons with respect to
$x=2E/Q$, the particle energy normalized to the
total energy $Q$, at $Q=34$ GeV and $Q=91$ GeV, are shown
in Fig. 3b and again agree  with the data. Parton
shower Monte Carlos are well known to describe these
data sucessfully: our repetition of this success
simply means that our novel parton-hadron conversion
mechanism is innocuous in this regard.
The comparisons in Fig. 3 do not indicate any clear preference
for one value of $B$ or $L_c$ over the other.

However, some indication may eventually be
drawn from Bose-Einstein correlations among produced hadrons.
The observed Bose enhancements in the pion spectra
are generally interpreted as reflecting the size and
coherence of the hadron emission region, which is directly related to
the choice of $B$ and thus $L_c$ in our model.
We have used the method of Sj\"ostrand \cite{sjo93} to simulate
the enhancement for same-sign charged pion pairs.
We see in Fig. 4a that the enhancements are indeed distinguishable
for $B^{1/4}=230$ MeV ($L_c=0.6$ $fm$) and
$B^{1/4}=180$ MeV ($L_c=0.8$ $fm$). Moreover, these enhancements are
essentially independent of $Q$, as can be seen from the ratios
of the enhancement factors plotted in Fig. 4b. We compare
our predictions with OPAL data in Fig. 4a. {\it Prima facie},
these are very consistent with our predictions for
$B^{1/4}=180$ MeV ($L_c=0.8$ $fm$). However, the unambiguous
definition of a preferred value of this input parameter must await a
systematic analysis of available data on Bose-Einstein
correlations in $e^+e^-\rightarrow hadrons$.
Nevertheless, we are pleased that our model appears to be able
to correlate sucessfully such diverse quantities as the bag constant $B$,
the characteristic microscopic scale $L_c$ for the
parton-hadron transition, and measured Bose-Einstein correlations.
\smallskip

In conclusion: we presented a novel approach to the
dynamics of parton-hadron conversion and confinement, based on
a kinetic multi-particle description
in real time and complete phase-space, which combines perturbative QCD
at high energies and an
effective field theory approach to hadrons at low energies.
As a test application that exhibits generic features, we have
considered the prototype reaction
$e^+e^- \rightarrow hadrons$.
The main results are: (i) the local conversion of partons to
hadronic clusters occurs very rapidly,
but the global time scale for the transition of all parts of the
system is comparatively long;
(ii) features of the perturbative parton evolution are projected
unscathed onto  hadron distributions;
(iii)
the Bose enhancement of same-sign pions
may provide a sensitive probe of the details of the space-time evolution.
Our approach may be extended to other applications involving parton
cascades and hadronization in different environments, including
deep inelastic lepton-nucleon or -nucleus scattering,
high energy hadron-hadron, hadron-nucleus, or
nucleus-nucleus collisions, and
the (non-equilibrium) dynamics of the QCD phase transition
at finite density and temperature.
The development of a systematic understanding of the hadron emission regions
in different processes appears within reach.

%\newpage

\bigskip
\medskip
\newpage

{\bf FIGURE CAPTIONS}
\bigskip

\noindent {\bf Figure 1:}
{\bf a)}
Time evolution of the kinetic pressures $P_{qg}$ of partons and $P_\chi$
of prehadronic clusters for $q\bar q$ initiated jet evolution with
total jet energy $Q=100$ GeV.
{\bf b)}
Total transverse momentum $p_\perp$ generated during the
time evolution of the system in the center-of-mass of the initial
two-jet.
\bigskip

\noindent {\bf Figure 2:}
{\bf a)}
Distribution of the cluster sizes of clusters formed from neighboring partons.
{\bf b)}
Associated cluster mass spectrum.
\bigskip

\noindent {\bf Figure 3:}
{\bf a)}
Resulting average charged multiplicity versus total energy
$Q$ in $e^+e^-$ annihilation events, in comparison with experimental data.
{\bf b)}
$x$-spectra of charged hadrons ($x=2E/Q$) with respect to the variable
$\ln(1/x)$ at $Q=34$ GeV and $Q=91$ GeV, confronted with measured disributions
at PEP and LEP.
\bigskip

\noindent {\bf Figure 4:}
{\bf a)}
Simulated Bose-Einstein enhancement $b_{L_c}(q)$ as a
function of pair mass $q$ of emitted same-sign pion pairs
for the two values of $L_c$.
{\bf b)}
Ratios of the enhancements $b_{0.6\,fm}(q)/b_{0.8\,fm}(q)$
for total jet energies $Q=34$ GeV and $Q=91$ GeV.
\bigskip

\vfill
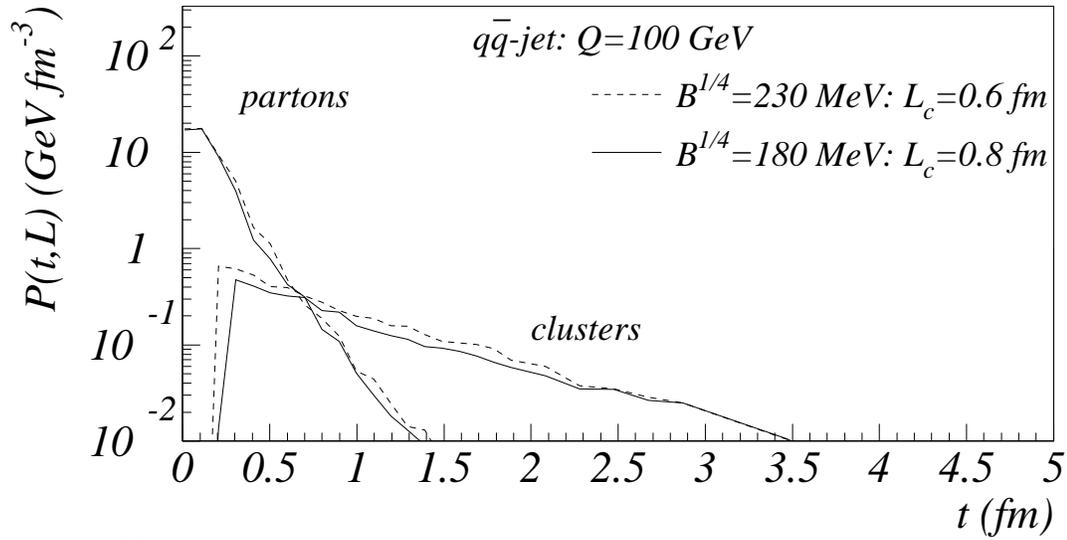
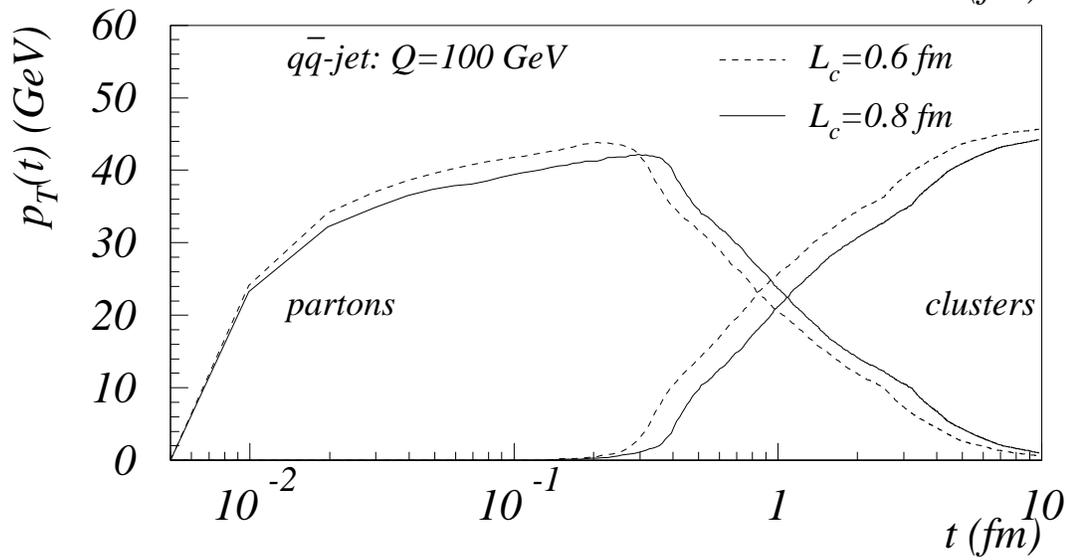
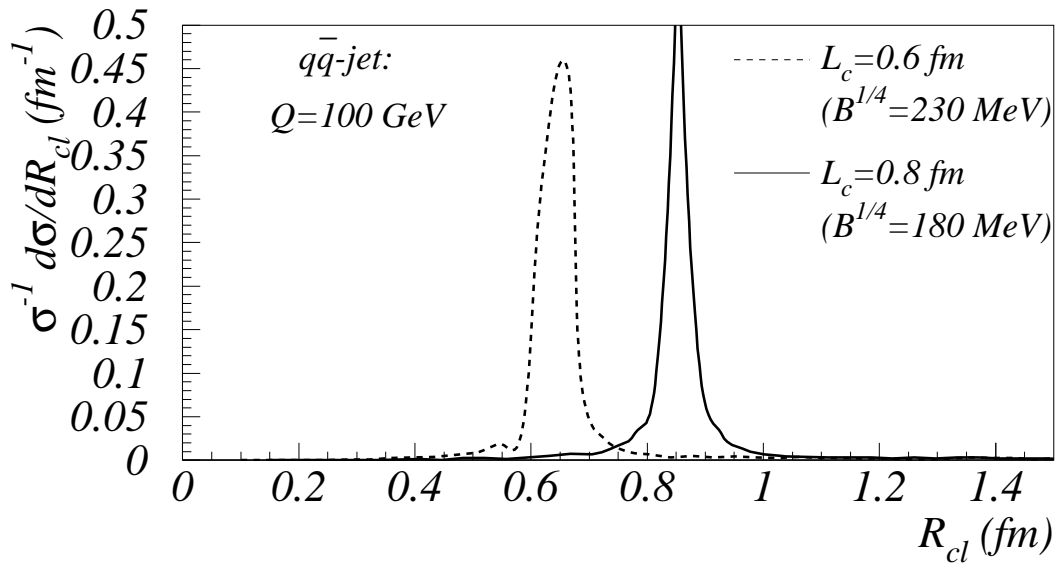
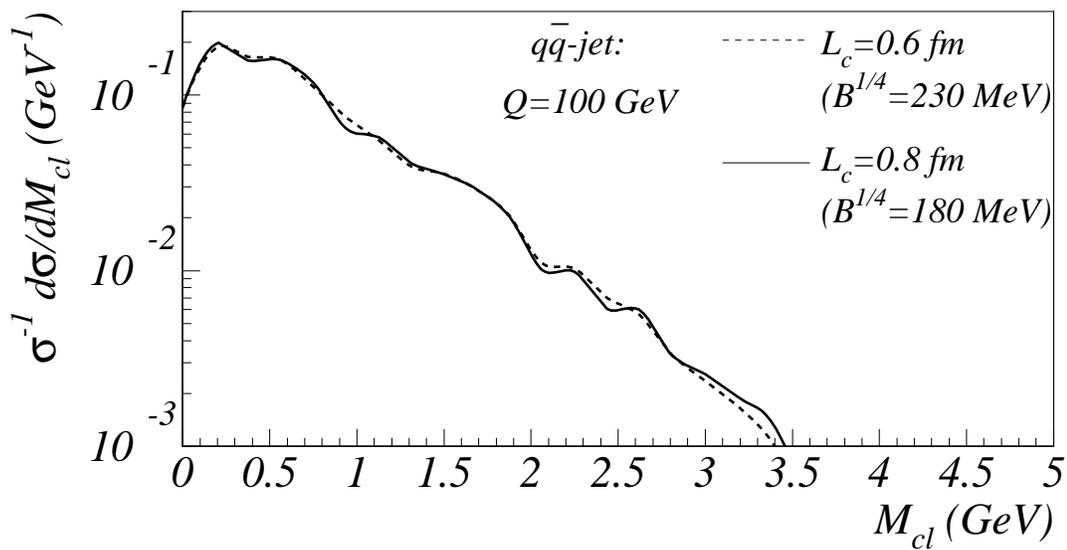
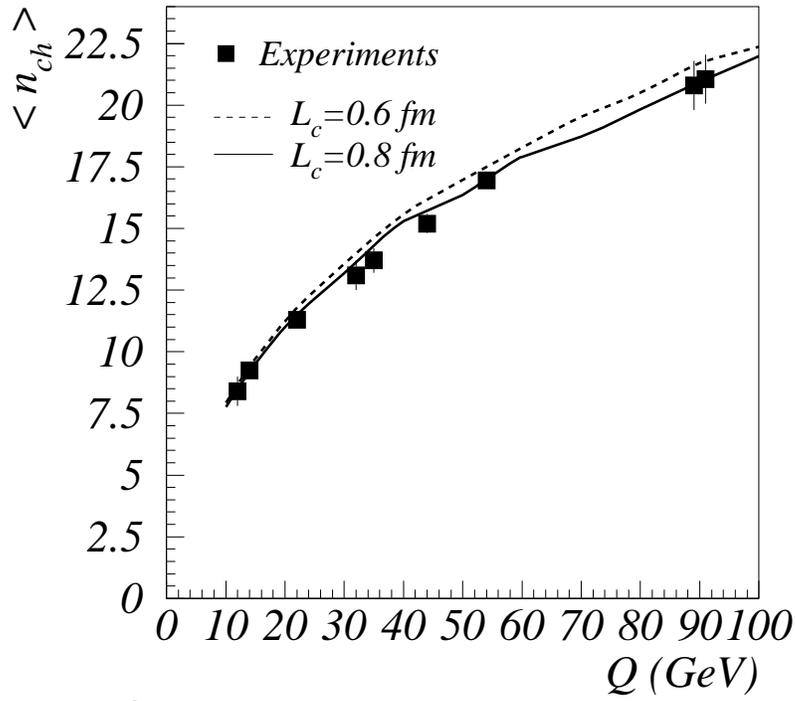
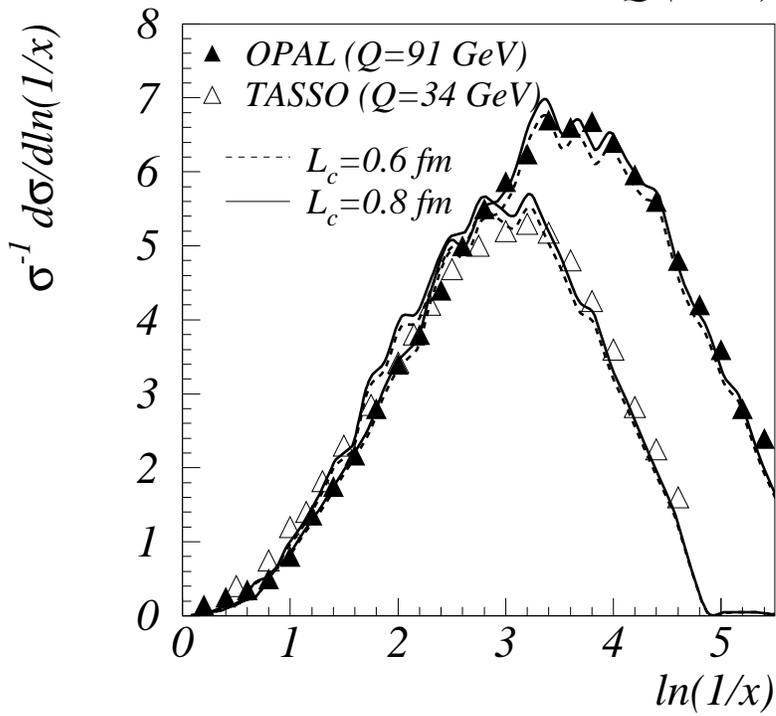
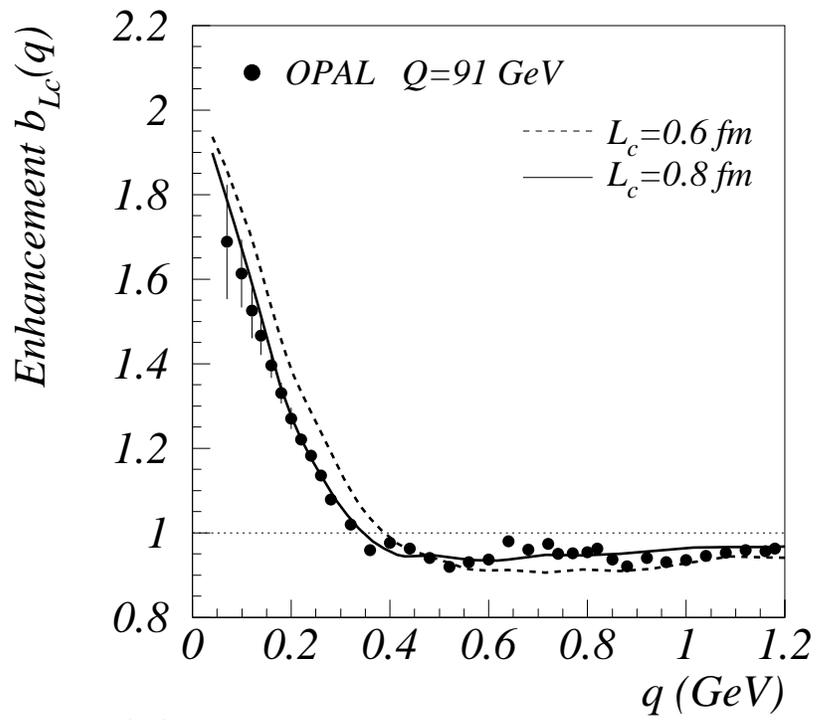
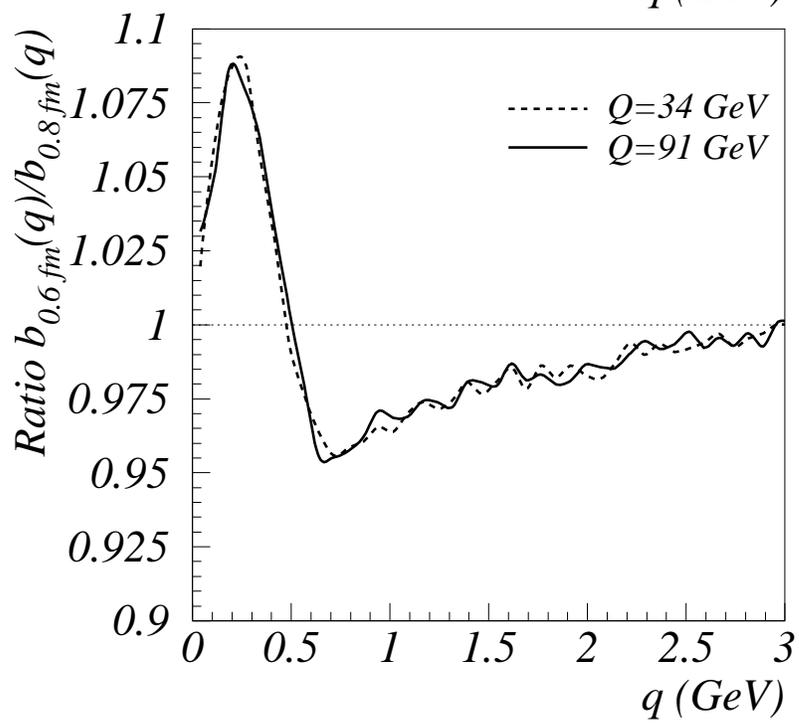
\end{document}